\documentclass[]{spie}  
\usepackage[]{graphicx}
\usepackage[]{amsmath}
\usepackage[]{url}

\title{The Extreme Polarimeter: Design, Performance, First Results \& Upgrades} 

\author{M. Rodenhuis\supit{a}, H. Canovas\supit{b}, S. V. Jeffers\supit{c}, M. de Juan Ovelar\supit{a}, L. Homs\supit{a}, M. Min\supit{d}, and C. U. Keller\supit{a}
\skiplinehalf
\supit{a}Sterrewacht Leiden, Niels Bohrweg 2, NL-2333CA Leiden, The Netherlands; \\
\supit{b}Departamento de Fisica y Astronomia, Universidad de Valpara\'iso, Valpara\'iso, Chile;
\supit{c}Institut f\"ur Astrophysik, Friedrich-Hund-Platz 1, D-37077 G\"ottingen, Germany;
\supit{d}Sterrekundig Instituut Anton Pannekoek, University of Amsterdam, Kruislaan 403, NL-1098SJ Amsterdam, The Netherlands
}

\authorinfo{Further author information: Send correspondence to M.R. E-mail: rodenhuis@strw.leidenuniv.nl, Telephone: +31 71 527 8432}

\begin{document} 
  \maketitle 

\begin{abstract}
Well over 700 exoplanets have been detected to date. Only a handful of these have been observed directly. Direct observation is extremely challenging due to the small separation and very large contrast involved. Imaging polarimetry offers a way to decrease the contrast between the unpolarized starlight and the light that has become linearly polarized after scattering by circumstellar material. This material can be the dust and debris found in circumstellar disks, but also the atmosphere or surface of an exoplanet.

We present the design, calibration approach, polarimetric performance and sample observation results of the Extreme Polarimeter, an imaging polarimeter for the study of circumstellar environments in scattered light at visible wavelengths.
   
The polarimeter uses the beam-exchange technique, in which the two orthogonal polarization states are imaged simultaneously and a polarization modulator swaps the polarization states of the two beams before the next image is taken. The instrument currently operates without the aid of Adaptive Optics. To reduce the effects of atmospheric seeing on the polarimetry, the images are taken at a frame rate of 35 fps, and large numbers of frames are combined to obtain the polarization images. 
   
Four successful observing runs have been performed using this instrument at the 4.2 m William Herschel Telescope on La Palma, targeting young stars with protoplanetary disks as well as evolved stars surrounded by dusty envelopes.  In terms of fractional polarization, the instrument sensitivity is better than 10$^{-4}$. The contrast achieved between the central star and the circumstellar source is of the order 10$^{-6}$. We show that our calibration approach yields absolute polarization errors below 1\%.
   
\end{abstract}



\section{INTRODUCTION}
\label{sec:intro}  

More than 700 exoplanets have been detected to date [\citenum{Schneider10}]. Almost all of these have been detected using indirect techniques such as radial velocity surveys and transit detections. The information that can be gathered about the planets from these detections is however limited. For the majority of the planets, detected through radial velocity measurements, only the orbital eccentricity and distance and a lower limit of the planetary mass can be obtained. Transits yield the planet radius, mass, orbital inclination and in a few cases the upper atmospheric composition [\citenum{Brown01, Charbonneau02, VidalMadjar03, VidalMadjar04, Deming05}]. Moreover, these two detection methods are subject to selection effects: radial velocity detections favor high-mass planets in close orbits. Transit detections are also biased towards short-period orbits and can only detect planets in orbital planes intersecting our line of sight to the star. 

To further characterize exoplanets we must be able to directly detect light from them and separate it from the light of the central star. Direct imaging of exoplanets is however a very challenging prospect. Not only is the separation between a star and its companion very small; the ratio of their intensities, the contrast, is extremely high. As a consequence, only a handful of exoplanets have so far been imaged directly (e.g. [\citenum{Lagrange09, Marois08}]).
These direct detections have all been performed in the (near) infrared, imaging warm, self-luminous planets around young stars. To observe and characterize cooler planets or even contemplate imaging of rocky planets, we require observations of starlight at visible wavelengths reflected by the planetary atmosphere or surface. Unpolarized light from the central star that is reflected by exoplanets becomes linearly polarized by the reflection process (e.g. [\citenum{Stam04}]). Imaging polarimetry is therefore an ideal way to discriminate between the polarized light from circumstellar environments and the unpolarized light from the nearby central star. The use of polarimetry to detect and study extra-solar planets has already been investigated by [\citenum{Seager00, Saar03, Hough03}].

We have built the Extreme Polarimeter (ExPo) with the aim of advancing this polarimetric technique. This imaging polarimeter has been developed with the goal of achieving an imaging polarimetry contrast better than 10$^{-5}$ unaided by Adaptive Optics. The first concepts for this instrument have been presented by [\citenum{Keller06}], the initial design by [\citenum{Rodenhuis08}]. Three successful observing runs have been performed using this instrument so far, all at the Nasmyth visiting focus at the 4.2 m William Herschel Telescope (WHT) on La Palma. The targets observed have included stars with circumstellar disks and protoplanetary disks or debris disks. In this paper we present the instrument design, polarimetric performance and sample results.

We present the measurement principle, beam-exchange polarimetry, in section 2. Section 3 describes the instrument design and section 4 the polarimetric calibration approach. Results from the first observing runs illustrating the instrument performance and scientific potential are presented in section 5. Our conclusions are presented in section 6.

\begin{table}
\caption{ExPo key parameters.}            
\label{table1}      
\centering                         
\begin{tabular}{l r}        
\hline
\noalign{\smallskip}
Wavelenght range (nm): & 500 - 900 \\
Field of view  ($''$ per beam): & 20$\times$30 \\
Image scale ($''$/pix): & 0.078 \\
Detector type: & Andor iXon EMCCD \\
Size (pixels): & 512$\times$512 \\
Pixel pitch ($\mu$m): & 16 \\
\noalign{\smallskip}
\hline                                 
\end{tabular}
\end{table}

\section{BEAM-EXCHANGE POLARIMETRY}
\label{sec:beamex}

Polarization measurements are inherently differential: In the case of linear polarization, two measurements taken through orthogonal polarization filters are subtracted. An unpolarized source will be equally present in both measurements and vanish in the difference, while a polarized source will remain. It is necessary to make the two measurements simultaneously, as temporal changes caused by atmospheric seeing will otherwise appear as false polarization signals. A polarizing beamsplitter can be used to do this. However, differential aberrations between the two beams will still appear as false signals. To solve this, a polarization modulator is used to alternately rotate the polarization by 90$^{\circ}$, effectively swapping the polarization information that is passed by the two beams. This so-called beam-exchange technique has been developed for sensitive solar polarimetry in the presence of atmospheric seeing [\citenum{Semel93}]. [\citenum{Keller96}] has shown this technique to limit errors to 2nd-order effects and to be able to achieve polarization sensitivities better than 1$\times$10$^{-5}$ of the total intensity. The technique is depicted schematically in Figure 1.

   \begin{figure}[h]
   \begin{center}
   \begin{tabular}{c}
   \includegraphics[width=16cm]{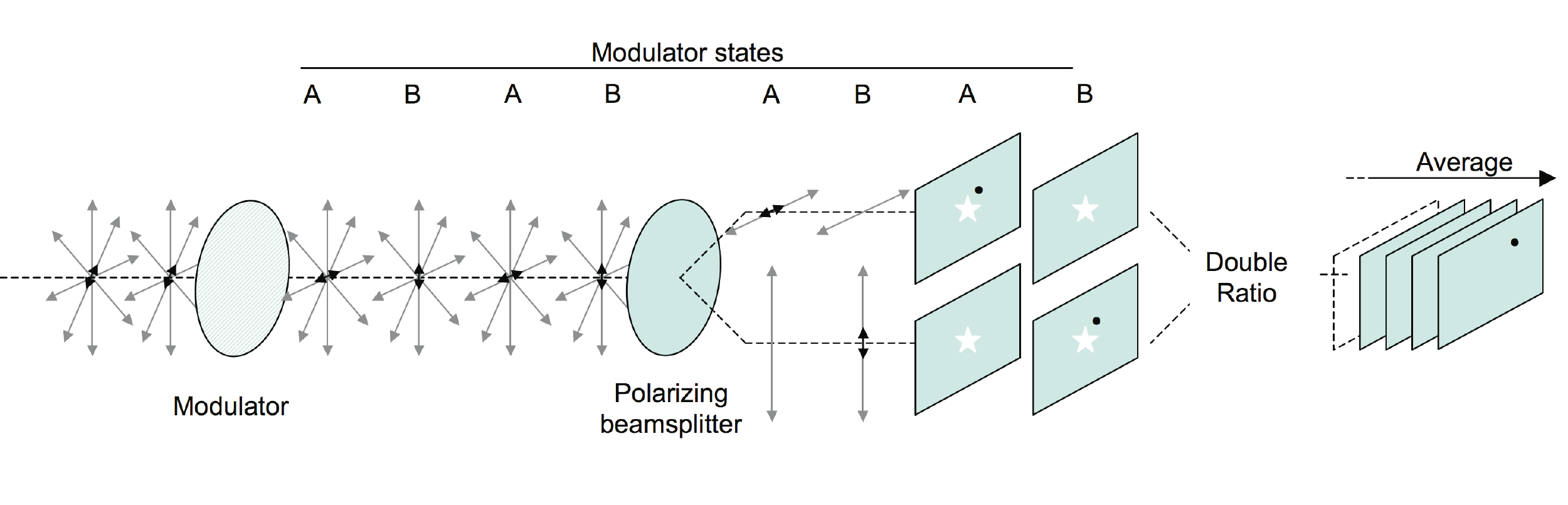}
   \end{tabular}
   \end{center}
   \caption[] 
   { \label{fig:beamexpol} 
Diagram demonstrating the dual beam-exchange technique being used to detect an exoplanet. The modulator alternately aligns the polarization orientation of interest with two orthogonal directions that are then separated by the polarizing beamsplitter. Both beams are imaged simultaneously by a camera. Subtracting the resulting images cancels out the large unpolarized signal while the (partially) polarized planet signal remains.}
   \end{figure} 

To reduce the effects of atmospheric seeing, we take long sequences of many short-exposure frames, combining the frames in sets of 4 and averaging over the sets. The short exposure times (typically 28 msec) ensures that the instantaneous PSF is relatively unchanged between the frames that are subtracted which would otherwise lead to false polarization signals. The alignment of frames within a set is performed based on the PSF, and may change slightly between one set and the next.
Using a polarization modulator has a further advantage: The filtering of the polarization state to which the instrument is sensitive takes place at the modulation element, making the instrument insensitive to instrument-induced polarization downstream of the modulator. It is thus desirable to place the modulator as far upstream in the instrument as possible.

\section{INSTRUMENT DESIGN}
\label{sec:design}
The ExPo design was created to perform five distinct but linked functions, each of which will be discussed in detail below:

\noindent \textbf{Polarimetry:} The core of the polarimeter consists of the polarization modulator and the analyzer. For dual-beam polarimetry, the analyzer needs to be a polarization beamsplitter. We have chosen a cube beamsplitter over crystal-based alternatives such as a Wollaston or (modified) Savart plate. The advantage of the cube is that the polarization beamsplitting is achromatic and does not suffer from crystal-induced astigmatism. The price we pay for this is the lower extinction ratio. Moreover, the extinction ratios is not the same for the reflected and the transmitted beams. ExPo currently uses a standard polarizing beamsplitter cube offered by Halle that provides an extinction ratios of 1:2000 and 1:40 for the transmitted/reflected beam over the 600-900 nm wavelength range.

The choice of polarization modulator has been driven by the fact that we wish to observe using short exposures in order to minimize the seeing variation during individual exposures and thus need to switch polarization states rapidly between exposures to achieve a high duty cycle. This means a mechanically rotating half-wave plate is not an option. We have chosen to use a Ferro-electric Liquid Crystal (FLC) device, which offers state transition times below 100 $\mu$sec. This is significantly faster than twisted nematic cells, for which the transition time is several milliseconds. Another difference is that with FLCs, the orientation of the fast axis is switched, while in nematics, it is the retardance that is modulated. A problem with FLCs (as with the nematics) is that by definition, the device is a half-wave retarder at a specific wavelength, and so is chromatic. It is possible to combine several waveplates in an achromatic combination [\citenum{Pancharatnam55}] and we intend to do this using three FLCs. However, other attempts to do this have not yielded encouraging results and so far we have used  a single FLC centered on 750nm.

The orientation of the FLC with respect to the polarizing beamsplitter determines the orthogonal polarization components (e.g. Stokes $\pm$Q) the instrument is most sensitive to, namely the the polarization angles that receive the maximum modulation. Polarization components at 45$^{\circ}$ to these angles (e.g. Stokes $\pm$U) will have minimal modulation, and thus the instrument will be insensitive to them. The FLC is therefor mounted on a rotation stage so the full linear polarization field can be measured.

ExPo has been designed to operate from a gravity-invariant position in the Nasmyth focus of the William Herschel Telescope. At this position, a significant amount of linear polarization is introduced by the 45$^{\circ}$ M3 reflection. The ExPo design includes a tilted glass plate to compensate this unwanted polarization. A control loop adjusts the tilt continuously after measuring the introduced polarization by comparing subsequent images taken at opposite modulator. The tilt assembly is mounted on a rotation stage that keeps the tilt axis aligned with the M3 orientation as the telescope elevation changes.
In order to perform polarimetric calibration, a linear polarizer can be inserted into the beam and rotated to any desired angle.

\noindent \textbf{Coronagraphy:} To reach high polarimetric sensitivity with a given dynamic range of the detector, it is desirable to suppress the peak of the stellar PSF. An added benefit is that the stray light in the instrument, and the possibility of it becoming polarized, is reduced. To this end, the ExPo design incorporates a coronagraph. We have opted for a classical Lyot-coronagraph consisting of a mask in the focal plane and a Lyot stop placed in the pupil. A large number of other coronagraph concepts exist but most tend to either introduce chromatic effects or be incompatible with polarimetry. Using a focal mask etched on a glass substrate allows us to use semi-transparent masks. The transmitted stellar peak can be used for precise guiding. The ExPo focal mask can be translated to place coronagraph dots of different sizes in front of the star.
The Lyot stop includes masking of diffraction by the telescope spiders, and is mounted on a rotator stage that keeps the mask aligned with changing telescope elevation. 

\noindent \textbf{Imaging:} In order to record both beams simultaneously and avoid synchronization issues associated with multiple detectors, ExPo relies on a single detector onto which both beams are projected. Two right-angle prisms mounted onto the polarizing beamsplitter are used to achieve this, as shown in Figure 2.

   \begin{figure}[h]
   \begin{center}
   \begin{tabular}{c}
   \includegraphics[height=7cm]{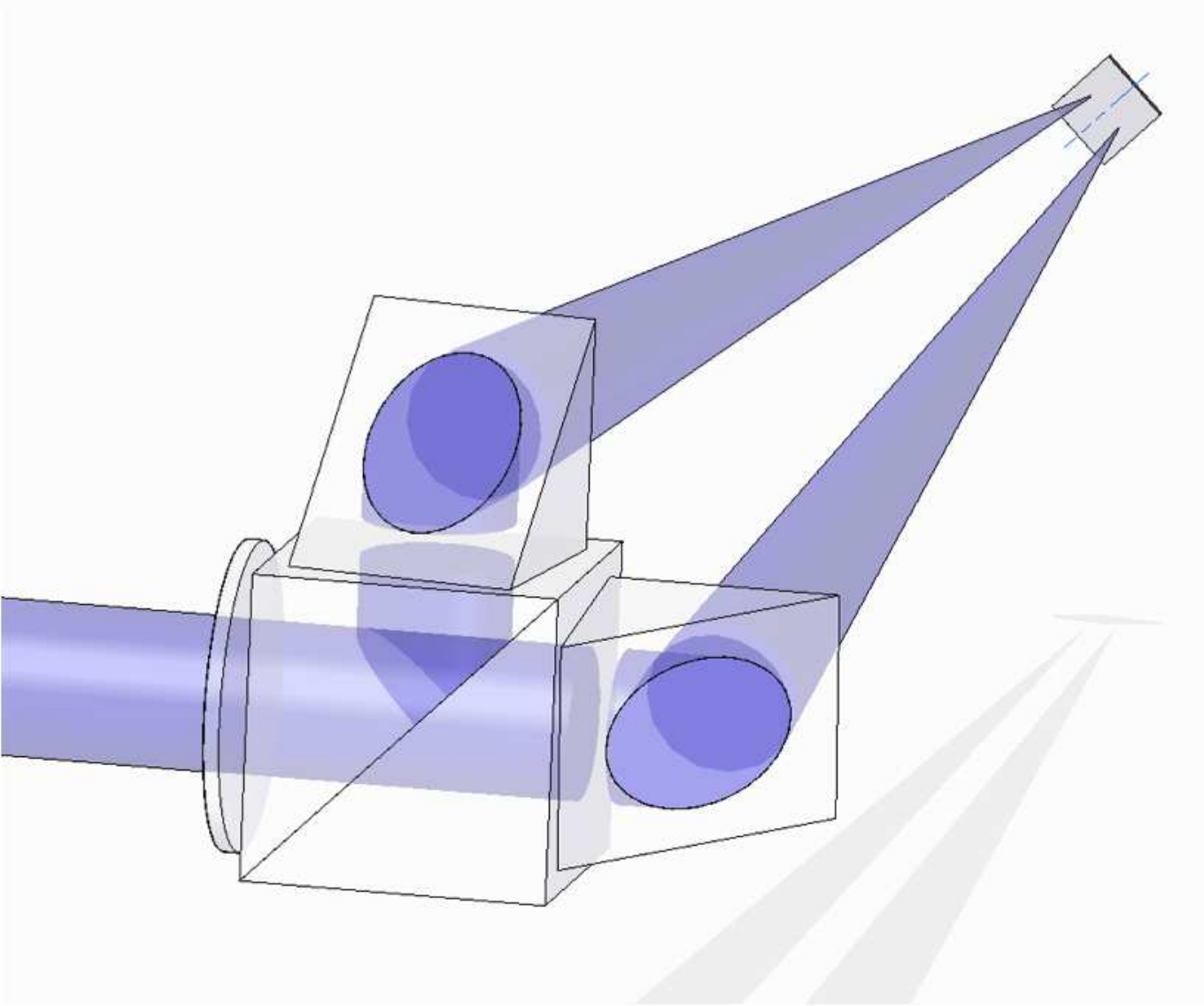}
   \end{tabular}
   \end{center}
   \caption 
   { \label{fig:prisms} 
After the beamsplitter, the beams are projected onto the detector by slightly rotating the two right-angle prisms.}
   \end{figure} 

The detector itself is rotated by 45$^{\circ}$ to align with the symmetry axis created by the beamsplitting cube. A disadvantage of this design is the slight projection angle introduced in the images. This effect is however symmetric; The images can still be combined.
As the instrument operates without the use of an Adaptive Optics (AO) system to correct for the atmospheric seeing, we wish to take large numbers of short exposures, with exposure times of the order of the speckle lifetime. Yet we also wish not to be limited to observing extremely bright targets by the detector readout noise. This has led to the selection of a detector using Electron-Multiplying CCD technology, where a signal can be amplified on the detector before being read out. The detector is an Andor iXon with a resolution of 512 $\times$ 512 pixels and a maximum framerate of 35 Hz. 
The design of the imaging optics also includes two fairly slow ($f/\#$ 10) lenses, with most optical elements placed in the collimated beam between them. The focal ratio demagnifies slightly to create a field of view of 20 $\times$ 30 arcseconds on each half of the detector. A smaller field of view can be obtained by increasing the focal length of the second lens. This lens is placed directly before the polarizing beamsplitter. 

\noindent \textbf{Atmospheric dispersion compensation:} The dispersion introduced by the atmosphere needs to be compensated before the coronagraph mask placed in the prime focus. A longitudinal ADC design has been chosen, primarily because any polarization introduced will be stationary in the telescope frame of reference. As with the polarization introduced by the Nasmyth flat, this can be compensated by the polarization compensator. The design consists of two wedge prisms, one fixed and one that can translate along the optical axis to increase the distance between the prisms, increasing the dispersion angle compensated. The complete ADC is mounted to the telescope derotator flange that can rotate to keep the prisms aligned with the plane of dispersion as the telescope elevation changes. For this design it was possible to use two identical off-the-shelf 3-inch prisms with a wedge angle of 3$^{\circ}$.   

\noindent \textbf{Wavelenght selection:} A filter wheel with 12 50 mm circular filter positions is used to perform ExPo observations with an array of broad- and narrowband filters.

\noindent Figure 3 presents the instrument block diagram below showing the integration of these five design elements.

   \begin{figure}[h]
   \begin{center}
   \begin{tabular}{c}
   \includegraphics[height=10cm]{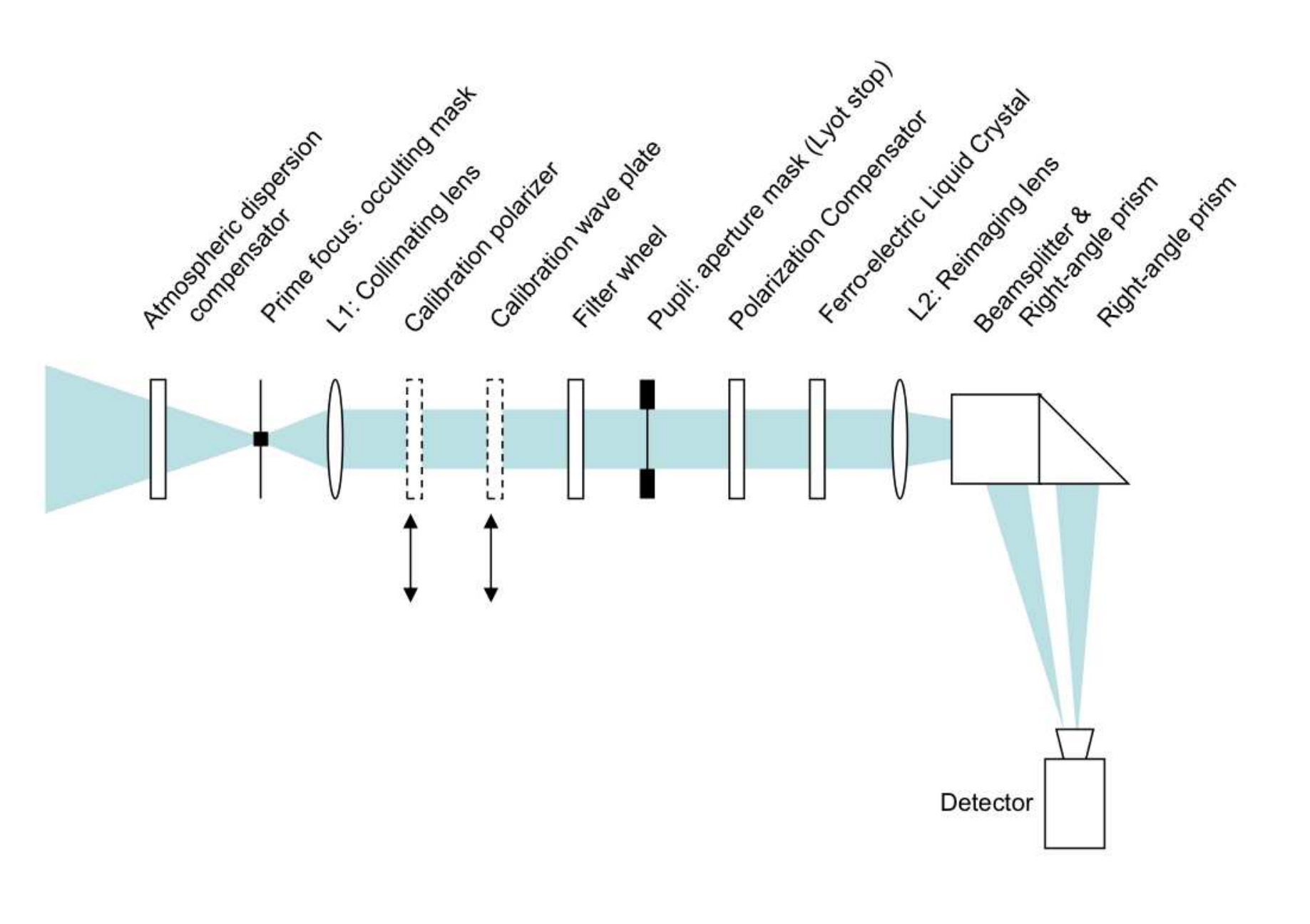}
   \end{tabular}
   \end{center}
   \caption
  { \label{fig:block} 
Schematic diagram showing the different ExPo components. The occulting mask in the prime focus and the aperture mask in the pupil image together form the coronagraph that suppresses the light from the star. The core of the polarimeter consists of the Ferro-electric Liquid Crystal (FLC) acting as the modulating element and the polarizing beamsplitter that splits the beam into its orthogonal polarization components.}
   \end{figure} 

In principle it is desirable to place the polarization modulator as far upstream in the instrument as possible. Any instrumental polarization introduced downstream will not be modulated and so cancel out in the differential imaging. The modulator however needs to be placed after the polarization compensator. As the compensatorÕs tilted glass plate introduces a slight and variable beam shift, it must be placed after the pupil mask, which requires precise alignment. This is the reason the modulator ends up relatively far downstream in the design.

Mechanically, the design is intended to be flexible, with components mounted and aligned individually on the optical table at the Nasmyth visiting instrument focus of the William Herschel Telescope. Much of the mounting hardware is off-the-shelf, with custom mounting hardware produced by our in-house workshop. After each observing run, the instrument is shipped back to our institute for testing and improvements in our optical lab. Figure 4 shows the instrument as mounted at the telescope.

   \begin{figure}[h]
   \begin{center}
   \begin{tabular}{c}
   \includegraphics[height=10cm]{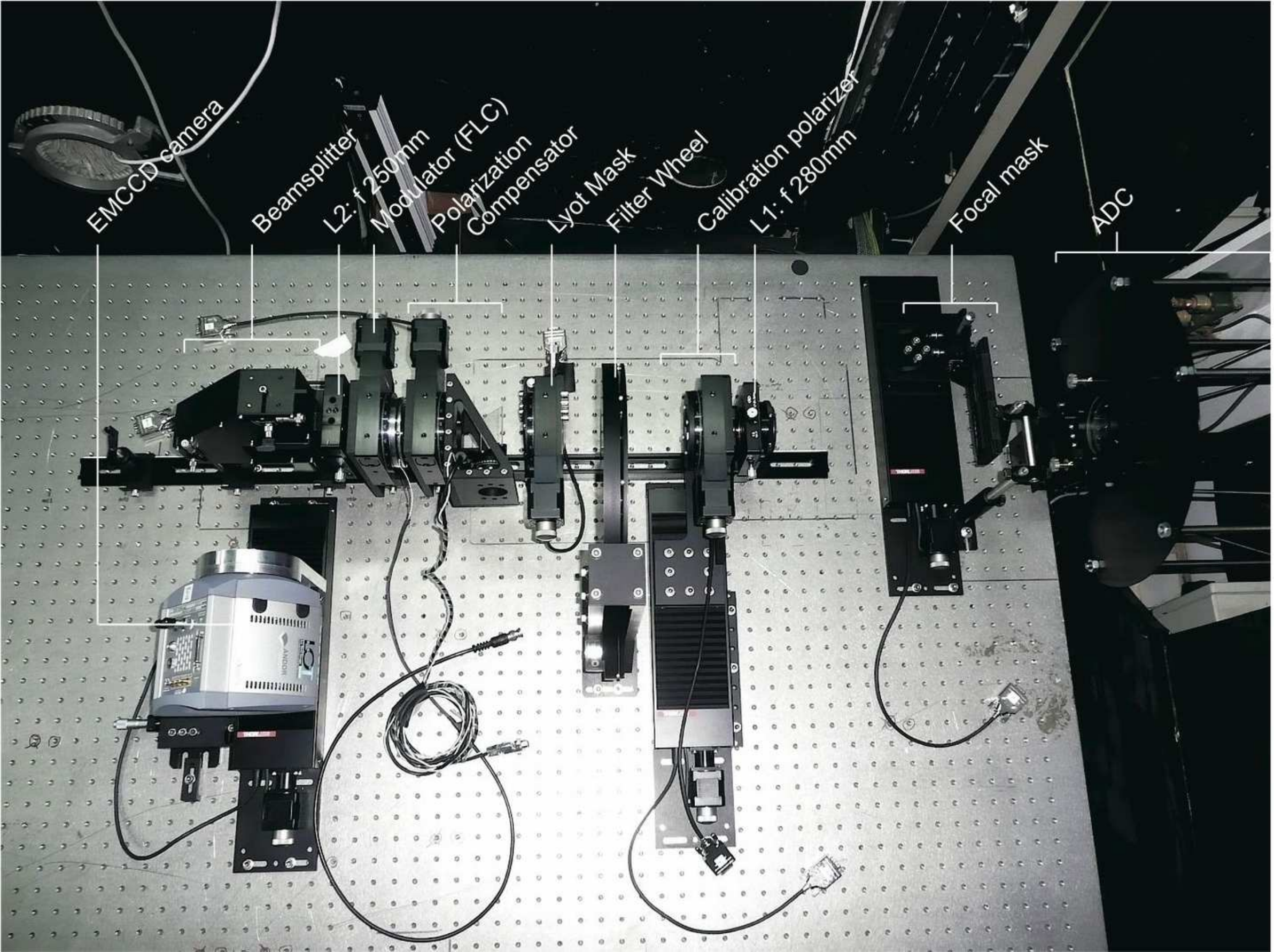}
   \end{tabular}
   \end{center}
   \caption 
   { \label{fig:expowht} 
Image showing the instrument as installed on the WHT Nasmyth port optical table during one of the observing runs.}
   \end{figure} 

\begin{table}
\caption{ExPo Photon budget (650 nm)}             
\label{table2}      
\centering   
\begin{tabular}{l c}  
\hline\hline
\noalign{\smallskip}
Element & Transmission \\     
\noalign{\smallskip}
\hline                       
  Atmosphere (V-band) & 0.90  \\
  Central obscuration & 0.94 \\
  Telescope primary & 0.90 \\
  Telescope secondary & 0.90 \\
  Nasmyth flat & 0.90 \\
  ADC & 0.98 \\
  Focal mask (no dot) & 0.95 \\
  Collimating lens& 0.98 \\
  Filter (Schott OG590 glass) & 0.90 \\
  Lyot stop & 0.75 \\
  Polarization compensator & 0.93 \\
  Polarization modulator (FLC) & 0.85 \\
  Refocussing lens & 0.98 \\
  Beamsplitter (reflected) & 0.48 \\
  Beamsplitter (transmitted) &0.42 \\
  Right-angle prisms & 0.98 \\
  CCD Quantum Efficiency & 0.93 \\
\noalign{\smallskip}
\hline
\noalign{\smallskip}
  Total (reflected beam) & 0.13 \\
  Total (transmitted beam) & 0.11 \\
\noalign{\smallskip}
\hline
\noalign{\smallskip}
  \multicolumn{2}{l}{Photons, $m_V = 0$ star:} \\
  Entering 4.2 m telescope: & 1.12$\times10^{11}$  ph/s \\
  Detected (reflected beam): & 1.4$\times10^{10}$  ph/s \\
  Detected (transmitted beam): & 1.3$\times10^{10}$  ph/s \\
\noalign{\smallskip}
\hline                                  
\end{tabular}
\end{table}

\section{DATA REDUCTION AND POLARIMETRIC CALIBRATION}
\label{sec:datared}

The raw product of ExPo measurements are image series in which the even and uneven frames correspond to the two different states of the FLC polarization modulator. Each raw image has a left and right sub-image formed by the two beams of the polarizing beamsplitter.

   \begin{figure}[h]
   \begin{center}
   \begin{tabular}{c}
   \includegraphics[height=7cm]{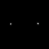}
   \end{tabular}
   \end{center}
   \caption 
   { \label{fig:rawframe} 
Raw ExPo frame showing the double image created by the two beams.}
   \end{figure} 

\noindent The FLC itself can be rotated to change the polarization direction to which the instrument is sensitive. The angle at which the instrument is sensitive to $\pm Q$ (in Stokes notation) is offset 22.5$^{\circ}$ from the angle for $\pm U$. For redundancy, ExPo observations are performed at four different FLC angles: 0$^{\circ}$, 22.5$^{\circ}$, 45$^{\circ}$, 67.5$^{\circ}$. With the combination of two FLC states (A, B) and the left/right sub-images, the beam-exchange technique thus yields four different types of images at every FLC angle. We denote these intensity images by:

For a general polarimeter, a minimum of four independent signals must be measured in order to determine the four elements of the Stokes vector. The signal matrix $\mathsf{X}$ relates the measured signals $\mathbf{S}$ to the input stokes vector $I$: 

\begin{equation}
\mathbf{S} = \mathsf{X} \cdot \mathbf{I}
\end{equation}

If $\mathsf{X}$ can be determined from a calibration source, its inverse can be used to obtain the Stokes elements for the observed targets:

\begin{equation}
\mathbf{I} = \mathsf{Y} \cdot \mathbf{S} \ \ \textup{with} \ \ \mathsf{Y} = \mathsf{X}^{-1}
\end{equation}

ExPo does not measure circular polarization, so we are only interested in the first three elements of the Stokes vector: $I$, $Q$ and $U$. For this, four measurements are more than sufficient. The two beams, two FLC states and four FLC angles combined provide 16 different measurements from which four must be chosen. However, choosing the A/B and left/right combination of images for a single FLC angle does not work. As the FLC exchanges the beams between the two modulation states, the A-left images will to a large degree contain the same polarization information as the B-right images. The same goes for the B-left and A-right pair, resulting in an ill-conditioned signal matrix $\mathsf{X}$. Another option is to combine all four FLC angles for a single choice of FLC state and beam, e.g. A-left. These four measurements are sufficiently independent and matrix $\mathsf{X}$ will be well-conditioned and invertable. But this approach results in another problem: As the observation at a single FLC angle may take up to an hour, observational conditions will have changed too much between the first and the last FLC angle for them to be combined in a meaningful way. Furthermore, as the telescope tracks the target across the sky, unpolarized artifacts such as ghosts may change position in the image and will then no longer vanish from the result.
To overcome these issues, we start by combining the images for a single FLC angle in the so-called double-difference:

\begin{equation}
S_{_{DD}} = \frac{1}{2} (S_{_{A,left}} - S_{_{A,right}} - S_{_{B,left}} + S_{_{B,right}})
\end{equation}

As the left and right sub-images are located on the same raw image, it is essential that they are carefully co-aligned before the above subtraction takes place. [\citenum{Canovas11}] discusses in detail the processing of raw images performed to generate the double-difference images.
Unpolarized signals drop out in this measurement, for which [\citenum{Keller96}] has shown that all first-order effects drop out. A reduced signal matrix now relates the two double-difference measurements for FLC angles 22.5$^{\circ}$ degrees apart to the linear polarization elements of the Stokes vector:

\begin{equation}
\left(
\begin{array}{ll}
S_{^{DD}}^{^{FLC1}} \\ 
S_{^{DD}}^{^{FLC2}}
\end{array}
\right) = \mathsf{X} \cdot
\left(
\begin{array}{ll}
Q \\ 
U
\end{array}
\right)
\end{equation}

The parameters of $\mathsf{X}$ are determined by taking dome flats with the calibration polarizer at different angles. A fitting routine is used to fit the response parameters a, b for each double-difference signal:

\begin{equation}
S_{_{DD}}  = 
\left( 
\begin{array}{ll} 
a & b 
\end{array}
\right) 
\cdot
\left(
\begin{array}{ll}
Q \\ 
U
\end{array}
\right)
\end{equation}
The initial fit assumes the light entering the calibration polarizer is unpolarized. In reality this is not the case: The dome lights may be polarized and the Nasmyth flat will certainly introduce polarization. We iterate the fitting by using the first fit to calibrate dome flats taken without the calibration polarizer and estimate the real input Stokes vector. A new fit is then performed using this estimate. Errors are estimated by back-calibrating the flats taken with the calibration polarizer. Two iterations are sufficient to reduce the errors to a few percent. For real observations, the instrumental polarization introduced upstream of the calibration polarizer is removed by the polarization compensator. Table 3 shows the increase in accuracy of the consecutive calibration iterations and the effect of the polarization compensator. 


\begin{table}
\caption{Dome flat polarization after each calibration iteration and the effect of the Polarization Compensator (PoCo).}
\label{table3}
\centering
\begin{tabular}{ccccc}
\hline\hline
\noalign{\smallskip}
Calib. & \multicolumn{2}{c}{PoCo Off} & \multicolumn{2}{c}{PoCo On}\\
iteration  &  Q (\%)  & U (\%) & Q (\%) & U (\%) \\
\noalign{\smallskip}
\hline
\noalign{\smallskip}
1 & 0.0 & 0.0 & 0.0 & 0.0 \\
2 & -0.02765 & 3.22021 & 0.00437 & 0.00243 \\
3 & -0.02740 & 3.21953 & 0.00437 & 0.00243 \\
4 & -0.02740 & 3.21953 & 0.00437 & 0.00243 \\
\noalign{\smallskip}
\hline
\end{tabular}
\end{table}

The current ExPo setup is quite chromatic, using only a single FLC and not an achromatic, Pancharatnam configuration. As we use dome-flats with a specific spectrum for our broad-band calibration, the calibration may not be correct for targets that have a substantially different spectrum. This is certainly the case for young stars peaking towards the blue, such as AB Aurigae, as the dome flat lamps are fairly red ($T_{eff}$ = 3000K). To overcome this problem, we have created a model of ExPo including the spectral response of all wavelenght-sensitive components, in particular the FLC, the polarization beamsplitter and the QE-curve of the detector. Using this model, we are able to generate synthetic calibration flats for any desired spectrum. The model has been calibrated to so that integrated dome flats generated with the dome flat spectrum are within 1\% of the real flats.

As unpolarized signals drop out in the double difference, Stokes $I$ cannot be determined from the double difference, but is instead found from:

\begin{equation}
I = \frac{1}{2} (S_{_{A,left}} + S_{_{A,right}} + S_{_{B,left}} + S_{_{B,right}})
\end{equation}

Note that this yields the calibrated intensity in polarimetry terms but does not take attenuation by the atmosphere and optical transmission losses into account. A photometric intensity calibration must be performed using standard stars.
The final step in the data reduction is the correction for sky rotation during the long observation sequences. This is achieved by first obtaining the double difference measurements from subsets corresponding to 30 seconds of observations. Each block of frames is physically rotated and calibrated while correcting for the rotation of polarization frame of reference. These blocks are then added together to obtain the final images in Stokes $I$, $Q$ and $U$.

\section{INSTRUMENT PERFORMANCE}
\label{sec:perf}

Three successful observation campaigns have been conducted using the ExPo instrument at the William Herschel Telescope on La Palma. A short fourth run was conducted with a smaller field of view of 10$'' \times15''$ , but this did not produce good results, partly due to adverse meteorological conditions. The observations have initially focused on stars known to or suspected of harboring protoplanetary disks as well as some debris disks. Here we present some results illustrating the performance of the instrument. Actual science results and analysis will be the topic of forthcoming papers.
The direction of the scattering polarization from a disk is perpendicular to the direction to the star. In the uncalibrated double-difference images, this polarization signature produces a double-lobed ÒbutterflyÓ pattern, illustrated in Figure 6.

   \begin{figure}[h]
   \begin{center}
   \begin{tabular}{c}
   \includegraphics[width=16cm]{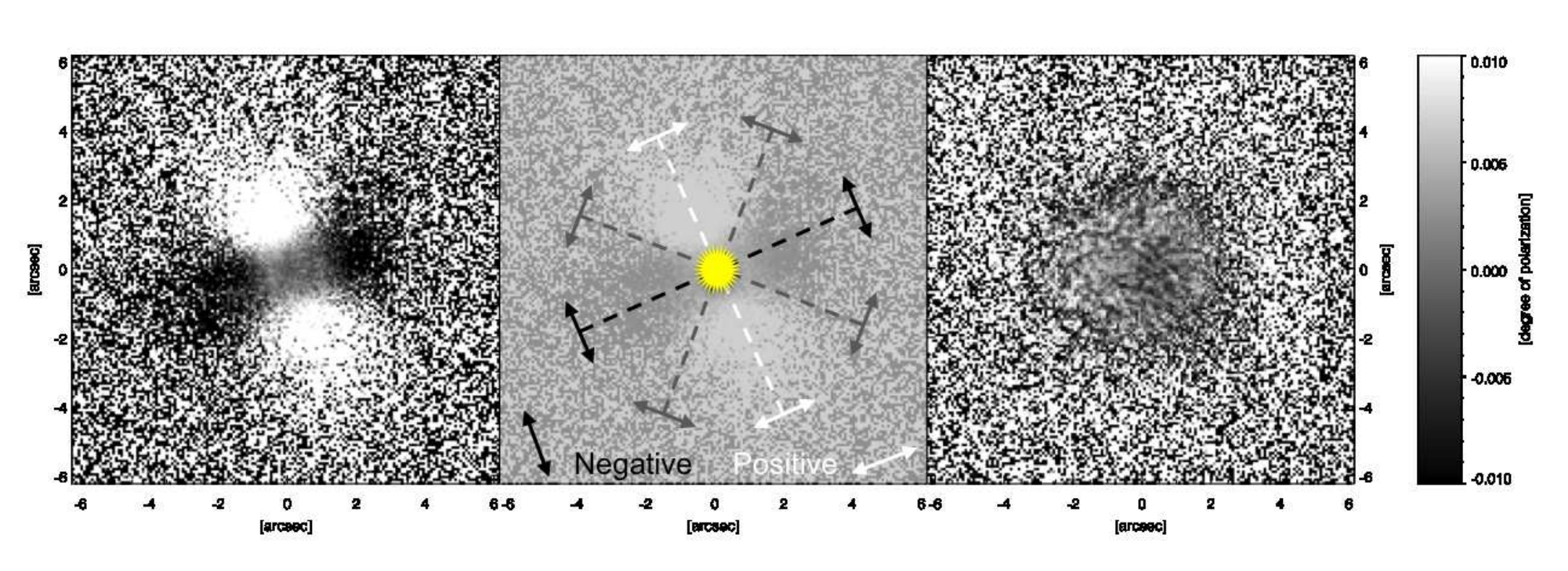}
   \end{tabular}
   \end{center}
   \caption 
   { \label{fig:butterfly} 
Uncalibrated fractional polarization images of the AB Aurigae protoplanetary disk illustrating the ÒbutterflyÓ signature produced by the tangential polarization pattern. For comparison, the image on the right shows the diskless star HD112815. The images are clipped at $\pm$1\% degree of polarization.}
   \end{figure} 

To demonstrate the advantage of the beam-exchange method, we have generated uncalibrated polarization images using only a subset of the data, emulating the performance of a single beam instrument and an instrument that does not include modulation. The results are shown in Figure 7.

   \begin{figure}[h]
   \begin{center}
   \begin{tabular}{c}
   \includegraphics[width=16cm]{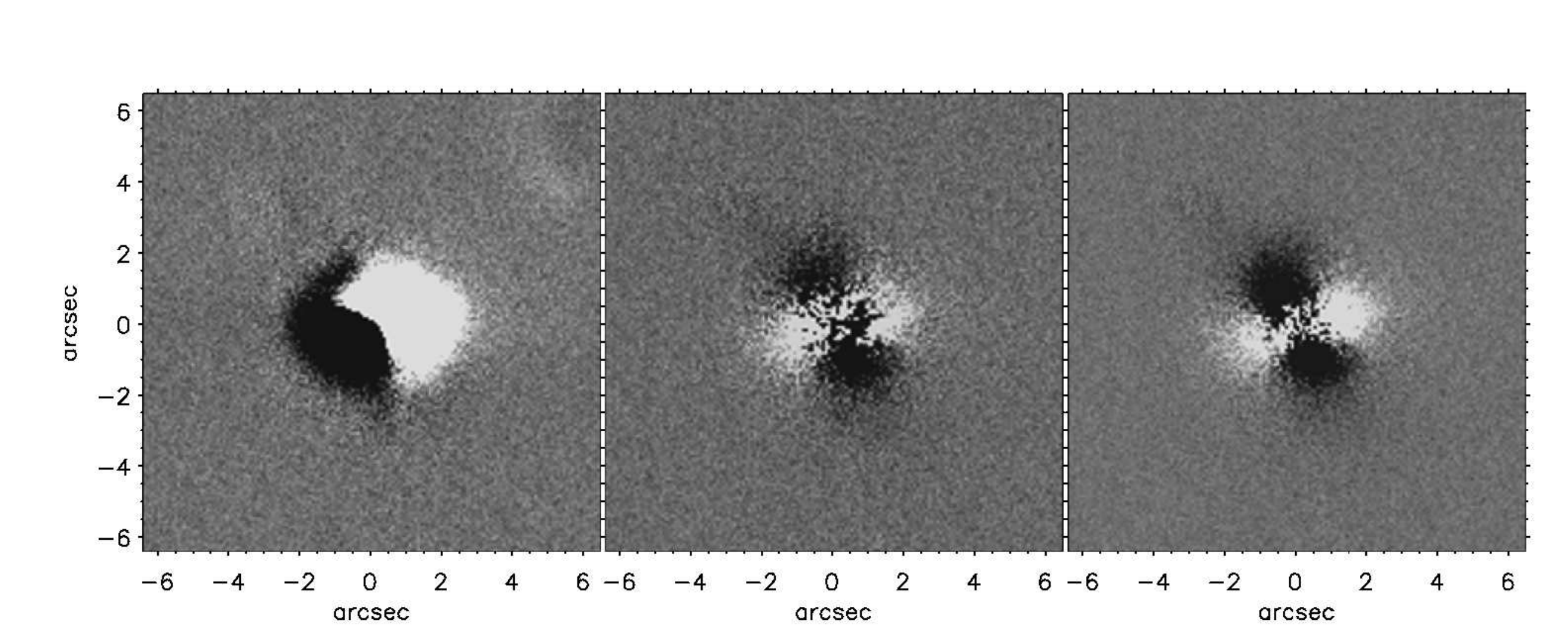}
   \end{tabular}
   \end{center}
   \caption 
   { \label{fig:abaurdualbeam} 
The advantage of the beam-exchange method. Left image: dual-beam without modulation. Middle: Single beam with modulation. Right: dual beam and modulation. Target is AB Aurigae.}
   \end{figure} 

In the left image, where the left and right beams are combined without modulation, differential aberrations between the two beams clearly destroy the butterfly pattern. In the middle image, in which subsequent modulated images from a single beam are used, the change in atmospheric seeing between the subsequent images produce smearing and a less defined butterfly pattern. The sharpest polarization pattern is clearly produced by the leftmost image, produced using the beam-exchange method.

The ExPo polarimeter is not intended primarily to provide extremely accurate absolute polarization measurements, but to be very sensitive to spatial variations in the polarization. Still, the calibration procedure described in the previous section yields results accurate down to the 1\% range. Table 4 presents typical accuracies when back-calibrating the dome flats taken at different calibration polarizer angles:


\begin{table}
\caption{Typical dome flat back-calibration residual calibration errors.}
\label{table4}
\centering
\begin{tabular}{ccc}
\hline\hline
\noalign{\smallskip}
Calibrator & \multicolumn{2}{c}{Using FLC 0 \& 22.5} \\
position & Error in $Q$ & Error in $U$ \\
\noalign{\smallskip}
\hline
\noalign{\smallskip}
$+Q$ & 0.17\% & 1.03\% \\
$+U$ & 0.22\% & 0.20\% \\
$-Q$ & 0.16\% & 1.00\%  \\
$-U$ & 0.22\% & 0.20\%  \\
\noalign{\smallskip}
\hline
\noalign{\smallskip}
  & \multicolumn{2}{c}{Using FLC 45 \& 67.5} \\
\noalign{\smallskip}
\hline
\noalign{\smallskip}
$+Q$ & 0.24\% & 0.63\% \\
$+U$ & 0.39\% & 0.34\% \\
$-Q$ & 0.22\% & 0.59\%  \\
$-U$ & 0.39\% & 0.34\%  \\
\noalign{\smallskip}
\hline
\end{tabular}
\end{table}

Observations of diskless and unpolarized standard stars must be performed in order to check that the instrumental polarization introduced by the Nasmyth flat is correctly compensated. Table 5 shows the calibrated Stokes vectors of two diskless stars, demonstrating that this is the case.


\begin{table}
\caption{Calibrated linear Stokes vector elements for two diskless stars. The values are averages of a 5 $\times$ 5 pixel area centered on the star.}
\label{table5}
\centering
\begin{tabular}{ccc}
\hline\hline
\noalign{\smallskip}
Calibration & \multicolumn{2}{c}{Using FLC 0 \& 22.5} \\
star &Measured $Q$ & Measured $U$ \\
\noalign{\smallskip}
\hline
\noalign{\smallskip}
HD86275 & -0.015\% & 0.002\% \\
HD107170 & -0.024\% & -0.023\% \\
\noalign{\smallskip}
\hline
\noalign{\smallskip}
  & \multicolumn{2}{c}{Using FLC 45 \& 67.5} \\
\noalign{\smallskip}
\hline
\noalign{\smallskip}
HD86275 & 0.047\% & -0.002\% \\
HD107170 & 0.014\% & 0.017\% \\
\noalign{\smallskip}
\hline
\end{tabular}
\end{table}
A more visual confirmation can be obtained by simply mapping the polarization of the central part of a diskless star. The result can be found in Figure 8.

   \begin{figure}[h]
   \begin{center}
   \begin{tabular}{c}
   \includegraphics[height=10cm]{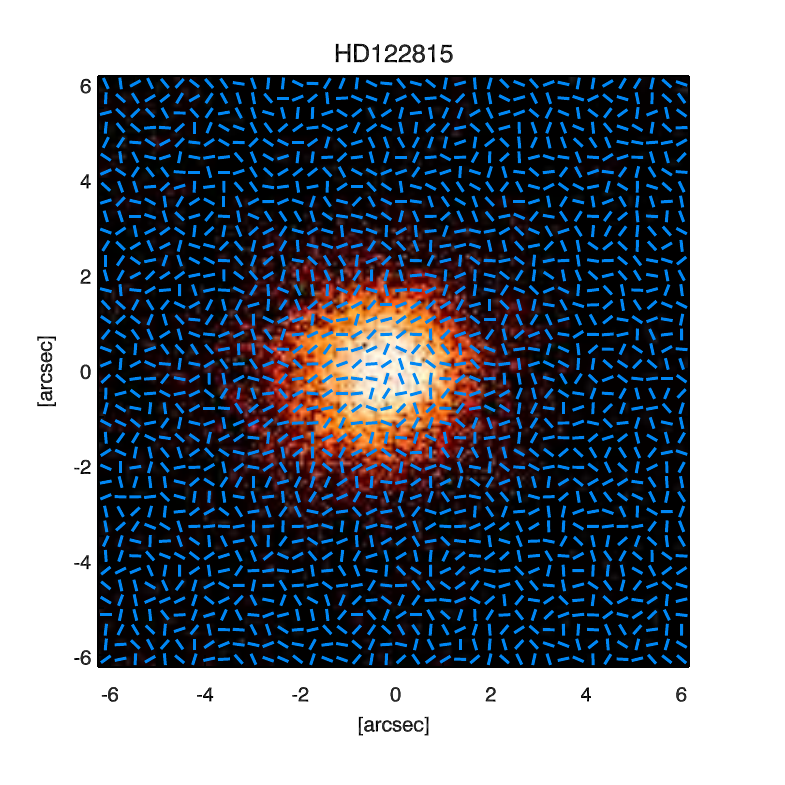}
   \end{tabular}
   \end{center}
   \caption 
   { \label{fig:unpol} 
Polarized intensity image of the unpolarized reference star HD122815 overlaid with calibrated polarization vectors. The reason we still see a signal is that the polarized intensity, defined as $\sqrt{Q^2+U^2}$, includes the absolute value of the photon noise. The vectors show no preferred orientation or other structure, showing that there is no residual instrumental polarization after the calibration.}
   \end{figure} 

A final analysis that can be performed using the diskless stars is to check how well the instrument rejects unpolarized light as a function of the radial distance to the star. Figures 9 and 10 show a radial profile of the achieved contrast and fractional polarization based on azimuthal averages. Figure 9 shows the total contrast: the polarized flux divided by the peak total stellar flux. The contrast achieved is 10$^{-4}$ at the star position, increasing to 10$^{-6}$ at about 4$''$ \ from the star. It should be noted that fewer frames were taken in the observation of this diskless star in comparison with regular science targets. The higher photon noise limits the contrast achieved. This is clarified by the Figure 10 showing the fractional polarization (polarized flux divided by total flux at the present radial distance from the star) plotted together with the profile of the photon noise detection limit. The measured fractional polarization closely traces this theoretical detection limit, except very close to the star, where the statistics become degenerate because of the small number of pixels that can be azimuthally averaged, and far away from the star, where there is hardly any signal detected and the result becomes very sensitive to small flat-field or dark correction errors.  The fractional polarization that can be detected is therefore limited by the photon noise. Higher contrasts can be achieved using longer exposure times.

   \begin{figure}[h]
   \begin{center}
   \begin{tabular}{c}
   \includegraphics[width=10cm]{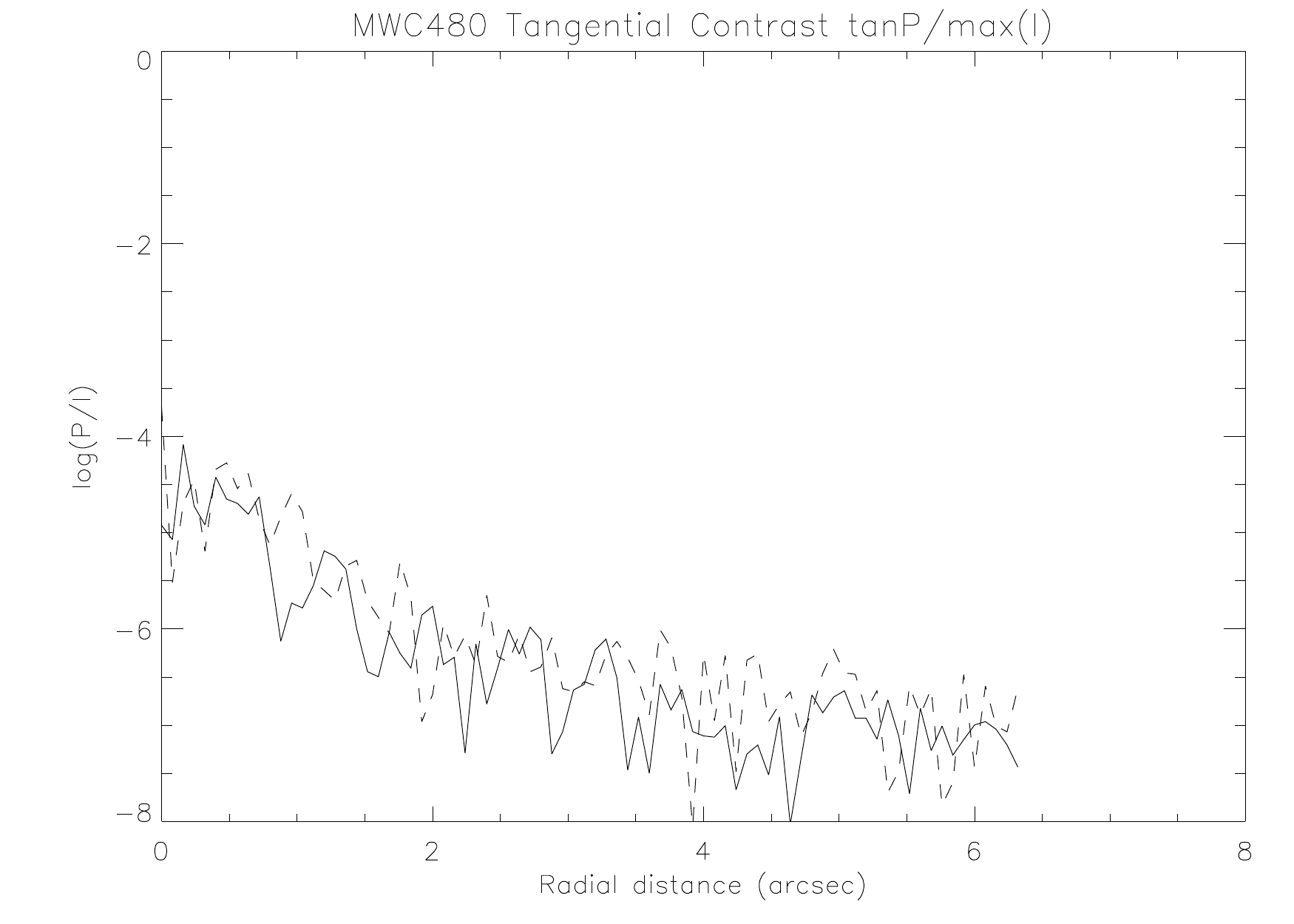}
   \end{tabular}
   \end{center}
   \caption 
   { \label{fig:contrast} 
Contrast reached for increasing distance from the star. At the peak of the star, the contrast reached purely by polarimetry is $10^{-4}$. This increases further out as the stellar intensity decreases. This target was observed for about 10 minutes, higher contrasts can be reached with longer observations.}
   \end{figure} 

   \begin{figure}[h]
   \begin{center}
   \begin{tabular}{c}
   \includegraphics[width=10cm]{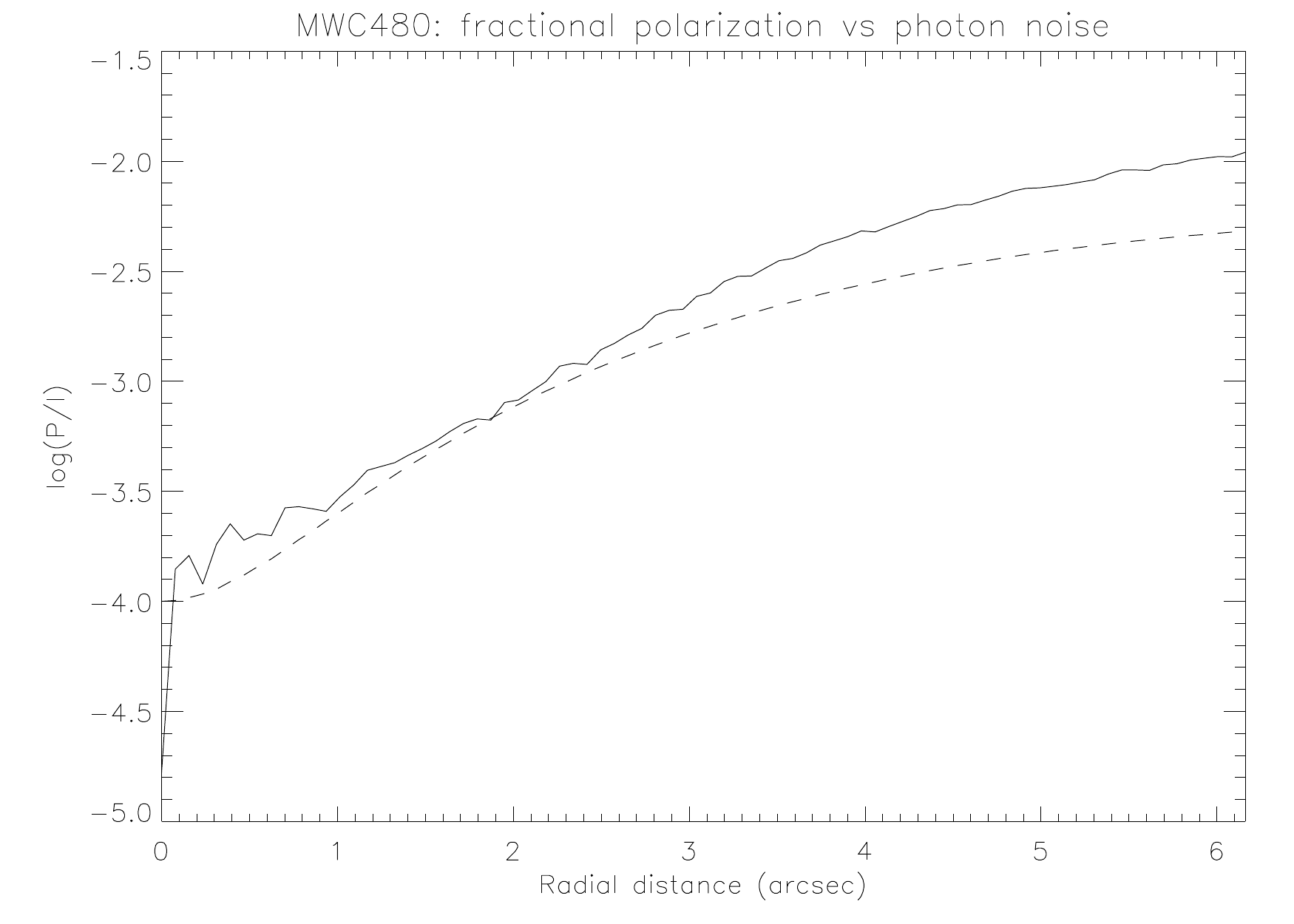}
   \end{tabular}
   \end{center}
   \caption 
   { \label{fig:fracpol} 
Fractional polarization averaged in circles around the star for increasing radial distance (solid line). The sensitivity is limited by photon noise and thus reaches its maximum close to the peak of the stellar flux. Dashed line: photon noise limit per pixel, averaged in circles around the star. }
   \end{figure} 

Having shown that our sensitivity is limited by the photon noise, it is clear that high-contrast polarimetry is a photon-hungry application. So while the polarization sensitivity and accuracy are key performance parameters for a polarimeter, another important characteristic is the efficiency with which these results are achieved, i.e. how well the instrument makes use of the available photons. [\citenum{delToroIniesta00}] define the polarimetric efficiency with which a polarimeter measures the Stokes elements as:

\begin{equation}
\varepsilon_{i} = \frac{1}{\sqrt{m \sum\limits_{j=1}^m \mathsf{Y}^{2}_{ij}}}
\end{equation}

With $\mathsf{Y}$ being the inverse of the signal matrix $\mathsf{X}$ as described in the previous chapter, $m$ the number of measurements and $i$ the index of the Stokes element. This expression makes no assumptions on how many measurements are performed to determine the Stokes element and how these measurements are performed. It is thus a global measure of the polarimeter efficiency. The efficiencies per Stokes element are defined so that:

\begin{equation}
\varepsilon_{Q}^{2}+\varepsilon_{U}^{2}+\varepsilon_{V}^{2} \leq 1
\end{equation}

As we do not measure circular polarization, the efficiency in V can be ignored. If we then require equal efficiency in Stokes Q and U we see that the maximum theoretical efficiency that can be reached is $\sqrt{0.5}$ = 0.707. Table \ref{table6} presents the efficiencies calculated for the individual beam and FLC state combinations (not the double difference) normalized with this theoretical maximum. For the individual beam / FLC state combinations, all four FLC angles are used to generate the calibration matrices. 


\begin{table}[!ht]
\caption{ExPo polarimetric efficiency per stokes element for individual FLC state \& beam combinations.}
\label{table6}
\centering
\begin{tabular}{cccc}
\hline
\noalign{\smallskip}
FLC state & Beam & \multicolumn{2}{c}{$\varepsilon_i$} \\
  &  &  Q  & U  \\
\noalign{\smallskip}
\hline
\noalign{\smallskip}
A & Left & 0.79 & 0.79 \\
  & Right & 0.81 & 0.82  \\
B & Left & 0.79 & 0.79  \\
  & Right & 0.79 & 0.81  \\
\noalign{\smallskip}
\hline
\end{tabular}
\end{table}

\section{SAMPLE RESULT: AB AURIGAE}
\label{sec:result}

As a sample science result, we present our imaging polarimetry observations of the well-known circumstellar disk around AB Aurigae. Imaging polarimetry is an excellent tool to observe the faint, linearly polarized light scattered by the circumstellar disk material while suppressing the unpolarized starlight. \citenum{Kuhn01} and \citenum{Apai04} have presented early studies of circumstellar disks using this technique. 

The protoplanetary disk around AB Aurigae, a 4$\pm$1 Myr old Herbig star, has been shown [\citenum{Grady99, Fukagawa04}] to have several spiral arms. Imaging polarimetry observations by [\citenum{Oppenheimer08}] revealed an annulus with a gap at about 100 AU from the star. This gap was speculated to be caused by a companion, but [\citenum{Perrin09}] showed that the structure can also be explained by the geometrical projection effect of the inclined disk. Only recently [\citenum{Hashimoto11}] presented very high resolution imaging polarimetry of the AB Aurigae disk. Their results reveal detailed structure in the inner region (up to 150 AU) of the disk, as well as clearly revealing the spiral arms extending out to about 450 AU. These results were all achieved using AO-assisted (or space-based in the case of [\citenum{Perrin09}]) coronagraphy in the near-infrared. 

Our observations date from January 2$^{nd}$ 2010. The imaging has been performed without the use of a color filter, thus spanning the full 400-900 nm wavelength range of our detector. Other observations have been performed using a Schott OG590 filter, but the results are inferior. This is not surprising given the A0e spectral type of AB Aurigae and the higher scattering efficiency towards the blue end of the spectrum. The average seeing during the observations was 1.0$''$. 

Figure 11 shows two calibrated polarized intensity images, each generated by combining two sets of observations with the position of the FLC modulator 22.5$^{\circ}$ apart. The structure in both images is very similar, even thought the seeing was slightly worse for the observations used in the right image. The disk is slightly elongated along the NE-SW axis, which is consistent with previous observations. In the inner region, up to about 1$''$\ from the star, we detect and elongated, almost bipolar structure oriented in the perpendicular direction. This structure is about at the highest resolution that can be resolved in our seeing-limited observations, but corresponds well to regions with higher polarized intensity to the NW and SE in the observations by [\citenum{Hashimoto11}]. Intriguingly, the first black contour in the left image appears to trace a similar annulus with a gap as reported by [\citenum{Oppenheimer08}], although the position is rotated slightly. Again, this is at the resolution limit and does not appear in the right-hand image, but both show traces of an annulus with a cleared inner region to the north of the star.

   \begin{figure}[h]
   \begin{center}
   \begin{tabular}{c}
   \includegraphics[width=15cm]{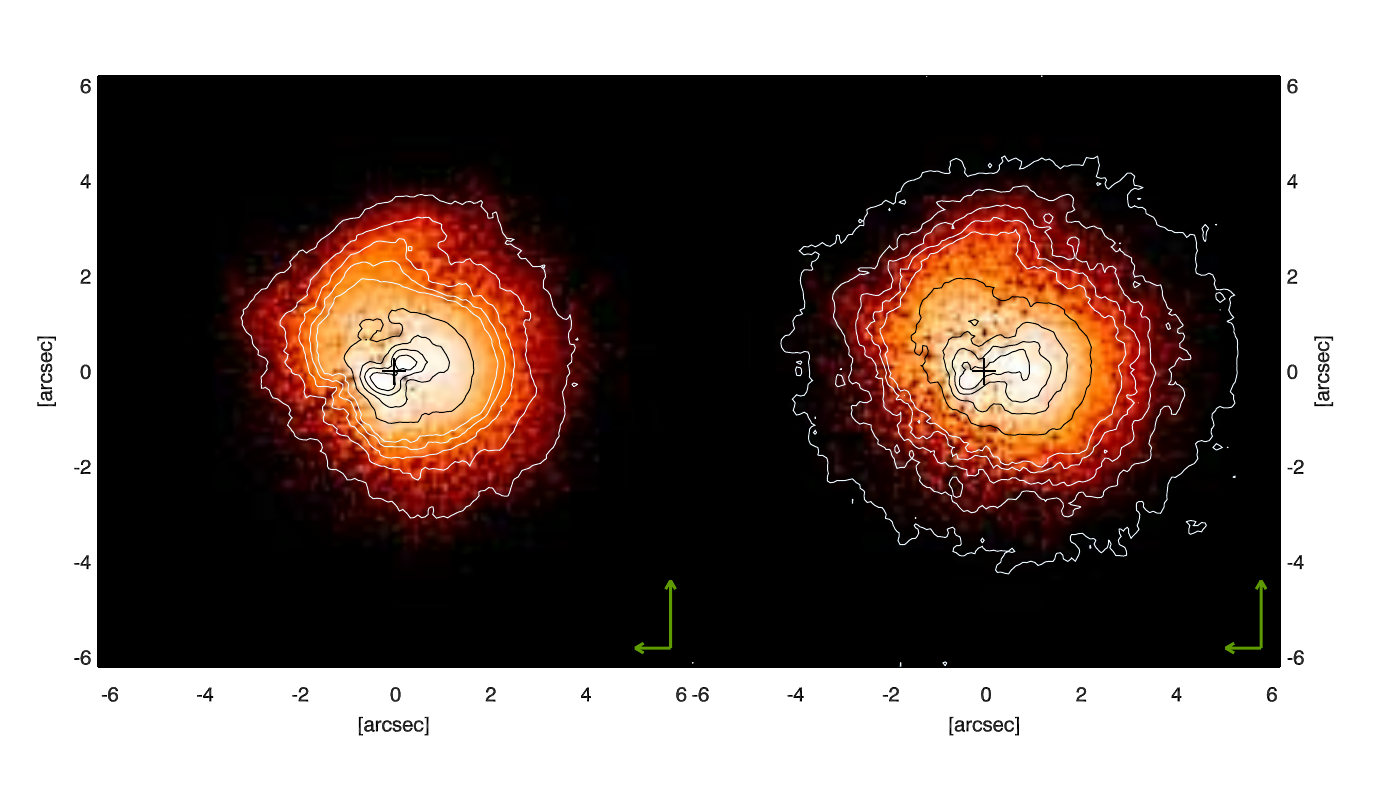}
   \end{tabular}
   \end{center}
   \caption 
   { \label{fig:abaupolint} 
Broadband-visible polarized intensity images of AB Aurigae. Observations at two positions of the liquid crystal polarization modulator have been combined to create each calibrated image:  0$^{\circ}$ and 22.5$^{\circ}$ for the left image, 45$^{\circ}$ and 67.5$^{\circ}$ for the right image. The color scale is logarithmic and histogram-equalized. Black contours go from 80\% of the maximum polarized intensity downwards in steps of 20\%, white contours go from 10\% of the maximum down in steps of 2\%.}
   \end{figure} 

What can be seen very clearly in both images is a spiral arm starting at about 150 AU (~1$''$) east of AB Aurigae and curving outward in the clockwise direction. To bring this structure out more clearly we have azimuthally averaged the image and then subtracted this from the original, producing Figure 12. The arm labeled 'S1' by [\citenum{Hashimoto11}] shows up very clearly, while there are hints of two other arms.

   \begin{figure}[h]
   \begin{center}
   \begin{tabular}{c}
   \includegraphics[width=15cm]{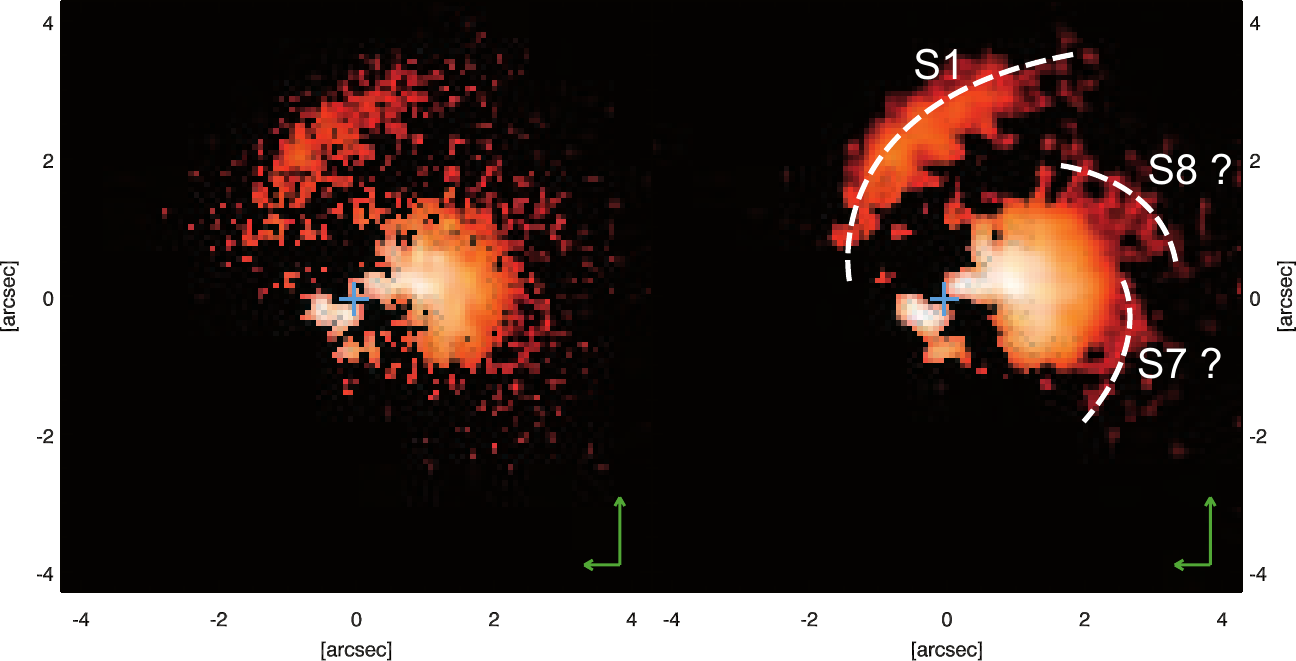}
   \end{tabular}
   \end{center}
   \caption 
   { \label{fig:abauazavsub} 
At left, the azimuthal average has been subtracted from the AB Aurigae polarized intensity image. At right, the image has been smoothed with a box-car width of 2 pixels. At least one of the spiral arms can clearly be identified. The arms have been labeled with the corresponding numbers from [\citenum{Hashimoto11}]. The color scale is logarithmic.}
   \end{figure} 

Figure 13 shows the disk polarized intensity overlaid with a vector plot mapping the polarization orientation. This is a useful test to ascertain that it is indeed the scattering polarization that is observed and that there are no offsets caused by errors in the calibration or the removal of instrumental and sky polarization. As expected, the polarization direction is perpendicular to the scattering plane, producing the azimuthal vector plot.

   \begin{figure}[ht!]
   \begin{center}
   \begin{tabular}{c}
   \includegraphics[width=10cm]{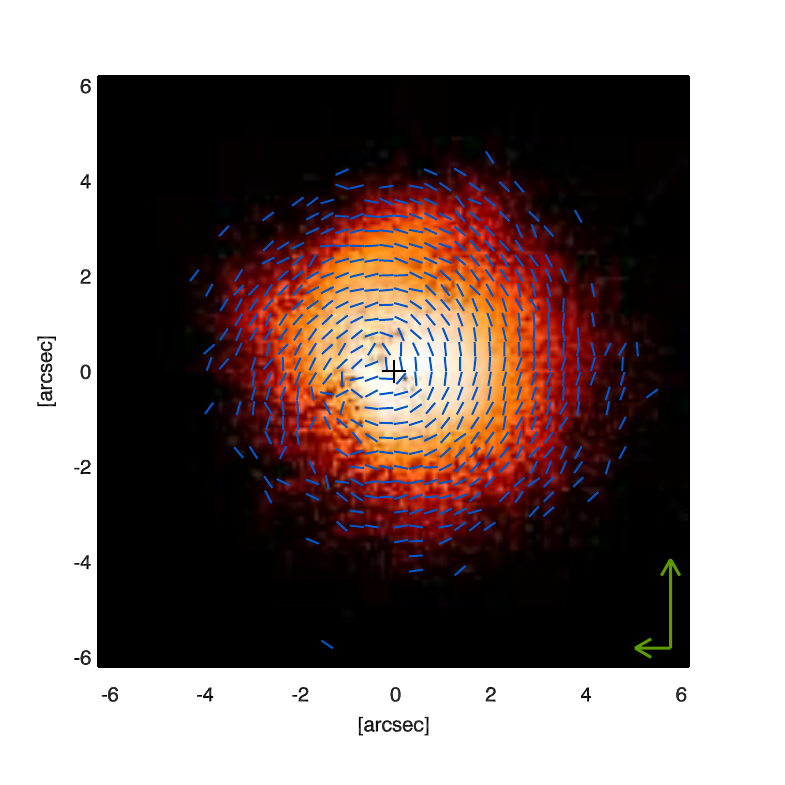}
   \end{tabular}
   \end{center}
   \caption 
   { \label{fig:abauvec} 
Polarized intensity image of AB Aurigae overlaid with a vector plot of the polarization direction clearly showing its azimuthal structure. }
   \end{figure} 

\section{EXPO UPGRADES}
\label{sec:upgrades}

An adaptive optics (AO) system has been developed for ExPo over the last year. This was done for a number of reasons: First of all, as the polarimetry is performed using large numbers of short-exposure frames, the increased image stability is expected to increase the polarization sensitivity significantly. The increased stability also means that the coronagraph built in to ExPo can now be used more effectively. Finally, the AO will hopefully increase the spatial resolution, and thus the level of detail we can see in the circumstellar structures. Simulations with the AO package PAOLA have shown that for 1$''$ seeing and a 12$^{th}$ magnitude star, an AO with a 50-actuator DM would improve the Strehl of the target by 20-25 times.

\begin{figure}[ht!]
   \begin{center}
   \begin{tabular}{c}
   \includegraphics[width=12cm]{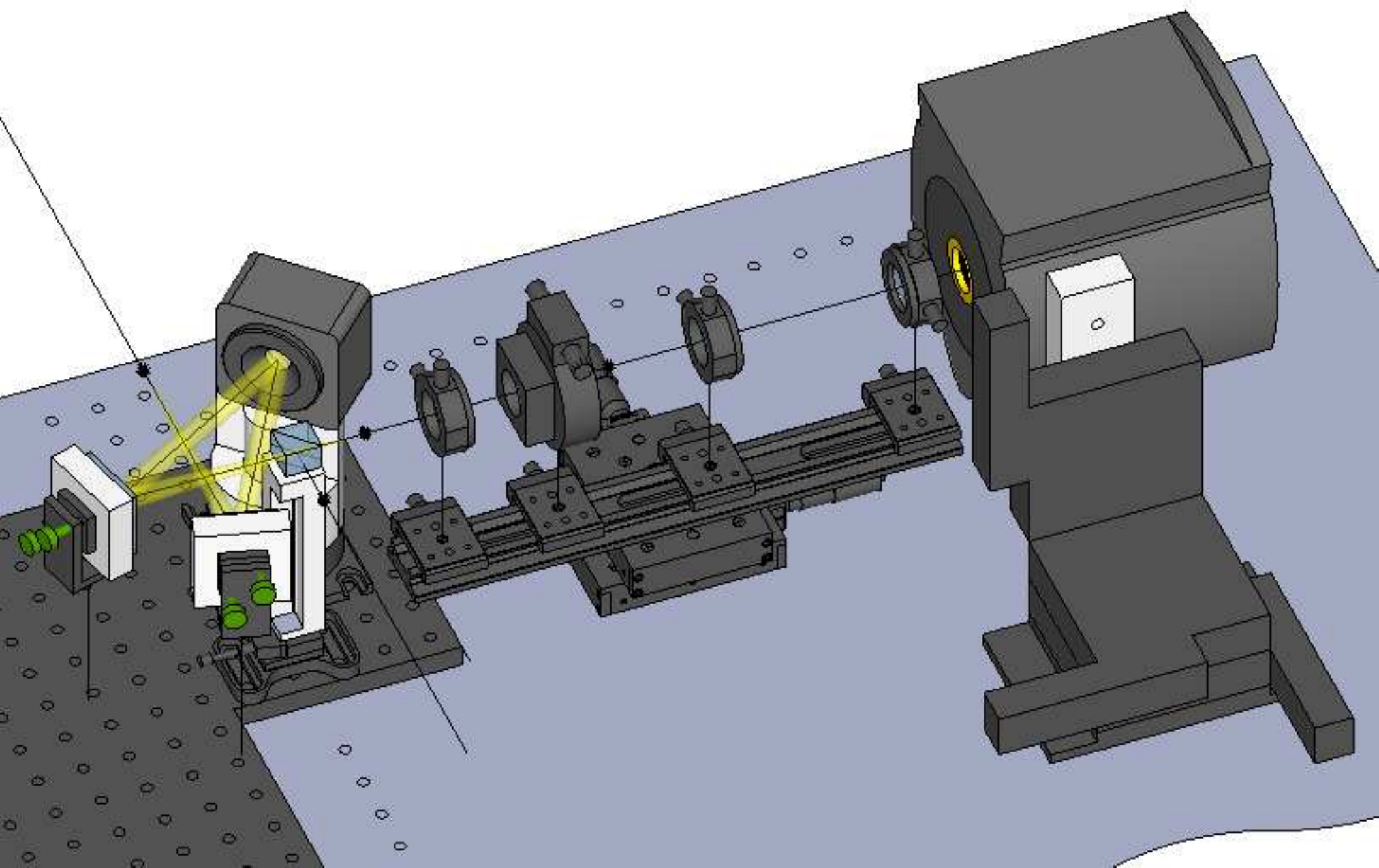}
   \end{tabular}
   \end{center}
   \caption 
   { \label{fig:expoao} 
CAD-drawing of the ExPo Adaptive Optics system. }
   \end{figure}

We have developed a custom design, intended to affect the polarimetry as little as possible. The adaptive optics system uses a 97-actuator deformable mirror (DM) from ALPAO. Two off-axis parabolas at reflection angles of 30$^{\circ}$ are used to form a pupil image on the DM and then form a secondary focus. Avoiding 45$^{\circ}$-reflections significantly decreases the instrumental polarization introduced. A non-polarizing beamsplitter is used to split the light between the science arm of the instrument and the Shack-Hartmann wavefront sensor. Figure \ref{fig:expoao} provides an impression of the optical and mechanical setup of the AO system.

The ExPo AO system had its successful commissioning run in the summer of 2012.

A second upgrade currently being worked on is a spectro-polarimetric integral field unit for the ExPo instrument. This is a technology demonstration project for the future EPICS-EPOL exoplanet imaging polarimeter for the E-ELT. The IFU will use a lenslet-based design to sample the image plane and generate low-resolution spectra for each sub-image. This will combine the high contrast provided by polarimetry with the ability to obtain spectral information of various points/areas of the target observed.

\section{DISCUSSION \& CONCLUSION}
\label{sec:conclusion}

The results we have presented demonstrate that our choice for the fast-modulating beam-exchange method and our relatively simple polarization-friendly optical design results in a sensitive imaging polarimeter, able to produce high-contrast observations of circumstellar environments without the aid of an AO system or a coronograph. While the sensitivity of around 10$^{-4}$ is not as high as some aperture-integrated polarimeters such as PlanetPol [\citenum{Hough06}] and POLISH [\citenum{Wiktorowicz08}], the spatial information is highly useful, not only because of its inherent potential to detect and study structures in circumstellar envelopes but also as a means to check the observations for global errors introduced by instrumental or sky polarization. The calibrated polarization accuracy of 1\% that can be reached with ExPo is similar to these instruments.

Being seeing-limited, the spatial resolution achieved with ExPo is not anywhere near that which can be achieved by current (e.g. HiCIAO: [\citenum{Hodapp08}]) and future (e.g. GPI: [\citenum{Perrin10}], SPHERE-ZIMPOL:  [\citenum{Thalmann08}]) AO-assisted coronographic imagers on 8-m telescopes. But none of these instruments have been designed specifically with polarimetry in mind and are likely to suffer sensitivity and accuracy losses incurred by the many reflections before the light arrives at the polarimeter module. Moreover, with the exception of Sphere-ZIMPOL, these instruments operate in the near-infrared, forgoing the inherently higher resolution and higher scattering efficiency available at shorter wavelengths.

With our sensitivity limited by photon noise, ExPo would of course benefit from operating at a bigger telescope or adding up more frames for a longer accumulated exposure time. The last option is however complicated by seeing variations that occur over the longer timescale. Apart from the obvious attraction of higher resolution, this is another motivation for developing a polarization-optimized AO system for ExPo. An AO system will also allow us to make better use of the coronograph included in ExPo. 

The suppression of starlight achieved by combining coronography and imaging polarimetry, coupled to the resolution and photon-flux provided by ELT-class telescopes and their associated extreme-AO systems, will enable contrasts of 10$^{-8}$ - 10$^{-9}$ to be achieved, the threshold at which Earth-like planets will be come detectable. We are currently applying the experience gained with ExPo to the design of the EPOL imaging polarimeter arm of the EPICS planet imaging \& characterization instrument for the European ELT.

\acknowledgments     
 
This research was funded through a VICI grant to Professor Keller from the Netherlands Organization for Scientific Research NWO.  


\bibliography{MRodenhuis_SPIE_8446_366}   
\bibliographystyle{spiebib}   

\end{document}